\documentstyle[prl,aps,multicol,epsfig]{revtex}
\begin{document}
\draft
\tighten

\title{ Quantum magnetism in the stripe phase: bond- versus site order.}
\author{Jakub Tworzyd\l o$^a$, Osman Y. Osman, Coen N. A. van Duin,
and Jan Zaanen}
\address{ Lorentz Institute, Leiden University, P.O.B. 9506, 2300 RA Leiden,
The Netherlands\\
$^a$ on leave from Institut of Theoretical Physics,
Warsaw University}
\date{\today ; jan@lorentz.leidenuniv.nl}
\maketitle

\begin{abstract}
It is argued that the spin dynamics in the charge-ordered stripe phase
might be revealing with regards to the nature of the anomalous spin
dynamics in cuprate superconductors. Specifically, if the stripes are
bond ordered much of the spin fluctuation will originate in the spin
sector itself, while site ordered stripes require the charge sector as
the driving force for the strong quantum spin fluctuations. 
\end{abstract}

\begin{multicols}{2}
\narrowtext
For quite some time it has been suspected\cite{Sa,SP} that the anomalous spin
dynamics of superconducting cuprates has to do with the $O(3)$ 
quantum non-linear sigma model (QNLS), describing the collective dynamics
of a quantum anti-ferromagnet\cite{CHN}. The discovery of the stripe 
phase\cite{tran} opens a new perspective on these matters. Below the
stripe-charge ordering temperature, charge fluctuations have to become
inconsequential and the remaining spin dynamics should fall automatically
in QNLS universality. As will be explained, the available data suggest
that this spin dynamics is characterized by a close proximity to the
QNLS zero temperature transition. This enhancement of the quantum-spin
fluctuations as compared to the half-filled antiferromagnet
can have a variety of microscopic sources. Here we will
focus on the possibility that these are due entirely to the charge-ordering
induced spatial anisotropy in the spin system. Although the influence
of spatial anisotropy is well understood on the field-theoretic 
level\cite{CNH,vDZa}, the charge can be bond-ordered or site ordered\cite{Za}
and this links the spin physics of the stripe phase to that of coupled spin 
ladders\cite{RiDa,Wa,AfflH}. At superconducting doping concentrations, bond- and
site order translate in coupled two-leg and three-leg spin ladders, 
respectively. We will present an in-depth quantitative analysis of both
problems, showing that spatial spin-anisotropy has to be largely irrelevant
for site-order, while it might well be the primary source of quantum spin
fluctuations in the bond-ordered case. A strategy will be presented to
disentangle these matters by experiment.

Let us first comment on the available information regarding the stripe phase
spin system. The spin ordering temperature appears
to be strongly surpressed as compared to half-filling\cite{tran}. 
A first cause can
be a decrease of the microscopic exchange interactions. However, the more
interesting possibility is that some microscopic disordering influence has
moved the antiferromagnet closer to the zero-temperature order-disorder
transition (quantum critical point). 
The few data available at present seem to favor the second possibility. 
We specifically refer to the ESR work by Kataev {\em et. al.}\cite{Kat} on 
$La_{1.99-x-y} Eu_{y} Gd_{0.01} Sr_{x} CuO_4$ exploiting the $Gd$ local
moments to probe the spin system in the $CuO$ planes. Quite remarkably,
little change is seen in the spin-lattice relaxation rate ($1 / T_1$) at
the charge ordering temperature, $T_{co} \simeq 70 K$. 
Above $T_{co}$ the $1/T_1$ 
is quite similar to that in $La_{2-x} Sr_x CuO_4$ where it is known 
from e.g. 
Neutron scattering that the magnetic correlation length $\xi$ is already
quite large at the temperatures of interest: since the width of the
incommensurate peaks is smaller than their separation, the correlation
length is larger than the stripe spacing\cite{Mason}. It follows that at 
$T \simeq T_{co}$ a continuum description of the spin dynamics should be
sensible. Below $T_{co}$ $1/T_1$ starts to increase exponentially upon
lowering temperature, signalling the diverging correlation length 
associated with the renormalized classical regime. Taken together,
this fits quite well the expectations for a quantum antiferromagnet 
which is rather
close to its quantum critical point, with a crossover temperature from
the renormalized classical- to the quantum critical regime $T^* \simeq T_{co}$.

\begin{figure}
\vspace{-\baselineskip}
\hspace{0.8cm}
\epsfig{file=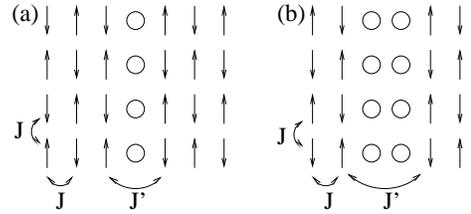,width=6cm}
\caption{Schematic distinction between site ordered (a) and bond ordered (b) 
stripes.}
\end{figure}

\vspace{-\baselineskip}
The increase of the coupling constant $g_0$, controlling the long wavelength
fluctuations, originates in some microscopic phenomenon. A limiting case
is that charge can be regarded as completely static even on the scale
of the lattice constant, such that its effect 
is to cause a spatially anisotropic distribution of exchange 
interactions\cite{CNH,vDZa}.
As indicated in Fig. 1, there are two options\cite{Za}: the stripes can be 
{\em bond} or {\em site} ordered.  
 It is expected that the spin dynamics associated with the
hole-rich regions is characterized by a short time scale and the magnetic
ordering phenomena are therefore associated with the magnetic domains. 
The spin-only model of relevance becomes either
a spin $S=1/2$ Heisenberg model
 describing three-leg ladders (site ordered) or two-leg ladders (bond
ordered) with uniform exchange interactions ($J$), 
mutually coupled by a weaker exchange interaction coupling ($\alpha J,
\alpha < 1$). This model is explicitely,
\begin{eqnarray}
 {\cal H} &=& J  \sum_{\vec{i}} 
            {\vec S_{ \vec{i}}} {\vec S_{\vec{i}+ \delta_y }} 
         + J \sum_{i_x \neq pn_{l}, i_y}
            {\vec S_{\vec{i}} } {\vec S_{ {\vec{i}}+{\delta_x} }} \nonumber\\ 
         & & + \alpha J
            \sum_{i_x=pn_{l},i_y }{\vec S_{ \vec{i}}} {\vec S_{{\vec{i}} 
          + {\delta_x}}},
\label{H-model}
\end{eqnarray}
where  $\vec{i}=(i_x,i_y)$ runs over a square lattice, ${\vec \delta_x}
= (1,0)$, ${\vec \delta_y} = (0,1)$. $n_l$ measures the width of the ladder
and $p$ counts the ladders. 

 Since the interest is in non-universal 
quantities as related to the non-trivial lattice cut-off, we studied the
model Eq.(\ref{H-model}) numerically using a highly      
efficient loop algorithm Quantum Monte-Carlo method\cite{ELM} supported with a
technique of improved estimators\cite{WY}. To keep track of the various
finite temperature cross-overs we focussed on the temperature dependence
of the staggered correlation length in both the directions parallel- ($\xi_y$)
and perpendicular ($\xi_x$) to the stripes. We typically insisted on
$3*10^4$ loop updates
for equilibration and $(2-3)*10^5$ updates for a measurement, keeping the
dimensions of the system in the $x$ and $y$ directions  
$L_{x,y}\geq 6*\xi_{x,y}$, to avoid finite size effects \cite{BBGW}.
The correlation lenght was determined by fitting the staggered spin-spin
correlation function $C({\vec r})  = (-1)^{r_x+r_y} \left< {\vec S_{i+r}}
 \cdot {\vec S_{ i}} \right>$, using a symmetrized
2 dimensional Ornstein-Zernike form $C(r)=A(r^{-1/2}e^{-r/\xi}+(L-r)^{-1/2}e^{-(L-r)/\xi})$ separately for
the $x$- ($\vec{r} = ( r, 0), L=L_x$) and $y$ (${\vec r}=(0,r), L=L_y$)
directions omitting the first few points to ensure asymptoticity. We
checked our results against the known results for both isolated ladders
by Greven {\em et. al.}\cite{GBW} ($\alpha=0$, $n_l=1,2,3$) and the
low temperature results for the isotropic ($\alpha =1$) limit\cite{WY,BW}.
 
Since $O(3)$ universality is bound to apply at scales much larger than
any lattice related cross-over scale, universal forms for the temperature
dependence of the correlation length can be used to further characterize
the long wavelength dynamics. The absolute lattice cut-off is reached at
a temperature ($T_{max}$) where the correlation length parallel to the 
stripes ($\xi_y$) becomes of order of the lattice constant. However, the 
problem is characterized by a second cut-off: when the correlation length
is less than the lattice constant in the direction perpendicular to the
stripes ($a_x$), the dynamics is that of $N_x$ independently
fluctuating spin ladders. We define $T_0$ as the temperature where 
$\xi_x \simeq a_x$ being the cross-over temperature below which the system
approaches 2+1D O(3) universality. In this latter regime, further cross-overs
are present. 
When the effective coupling constant ($g_0$) is less than the
critical coupling constant ($g_c$) a cross-over occurs from a `high'
temperature quantum critical-(QC) to a low temperature renormalized classical 
(RC) regime. In the QC regime $\xi \sim 1/T$ while the
cross-over temperature $T^*$ to the RC regime 
can be deduced from the exponential increase of 
the correlation length at low $T$, using\cite{CHN,BBGW,HN},
\begin{equation}
\xi(T)\propto \frac{e^{T^*/T}}{2T^*+T},
\end{equation}
where $T^*=2 \pi \rho_s$ in terms of the spin stiffness $\rho_s(\alpha)$.
When $g_0 > g_c$, the ground state is
quantum disordered (QD) as signalled by $\xi$ becoming temperature independent,
and the crossover temperature $T'$ between the QC- and QD regimes 
is estimated from the approximate relation\cite{GBW},
\begin{equation}
T'=\frac{c_y}{\xi_y(T\rightarrow 0) },
\end{equation}
where $c_y$ is spin wave velocity in the strong direction.

We determined the various cross-over lines as function of $\alpha$ for the
cases $n_l = 1, 2$ and $3$ (anisotropic Heisenberg, coupled two-
and three leg ladders, respectively). To determine $T_0$, 
we used for $\alpha$ close to 1
the same criterium as for the $T_{max}$ determination in the
isotropic problem  ($\xi_x(T^0)=0.7-0.8$). This
becomes inconsistent for small $\alpha$ where one better incorporates
the width of the ladder ($\xi_x(T_0)=n_l \times (0.7-0.8)$) and we
used a linear interpolation to connect smoothly both limits.
We checked that below the $T_0$, determined in this way,
both $\xi_x$ and $\xi_y$ exhibited the same dependence 
on temperature after an overall change of scale, 
demonstrating that the collective dynamics is indeed in
a 2+1D regime.    

\begin{figure}
\vspace{-\baselineskip}
\hspace{2mm}
\epsfig{figure=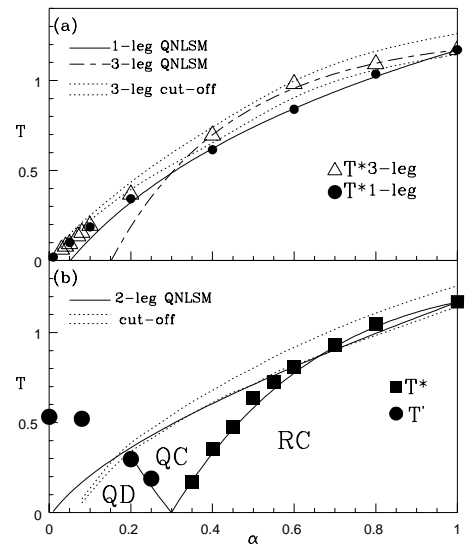, width=8cm}
\caption{Crossover temperatures as a function of
anisotropy $\alpha$ for the coupled three leg (a) and two leg (b) 
spin-ladder models. The lines and points refer to the analytical-
and numerical results, respectively, for the various scales.
Notice that the 1-leg `cut-off' (1D-2D cross-over) follows closely the
results for $T^*$.}
\end{figure}
 
\vspace{-\baselineskip}
In Fig. 2 we summarize our results in the form of a cross-over diagram as
function of $\alpha$ and temperature, both for the 1- and 3-leg (Fig. 2a)
and the 2-leg (Fig. 2b) cases. 
Consistent with analytic predictions\cite{Wa},
the behavior is radically different for the half-integer spin 1- and 3 leg
cases on the one hand,
and the `integer spin' 2-leg case on the other hand. 
Let us first discuss the former.
Here the ground state remains
in the renormalized classical regime for any finite $\alpha$. The reason
is obvious. In isolated ladders ($\alpha = 0$) with an uneven number of
legs the ground state is a Luttinger liquid exhibiting algebraic long 
range order and any finite ladder-to-ladder interaction will suffice 
to stabilize true long range order at $T=0$\cite{Wa,AfflH}. 
This in turn implies 
a finite $T^*$ where the classical nature of the ground state becomes visible.
Interestingly, our calculations indicate that $T^*$ and $T^0$ basically
coincide for any $\alpha$: at the moment the system
discovers that it is 2+1 dimensional, the classical behavior sets in. 
Our finding that $T^0$ increases
linearly with $\alpha$ for small $\alpha$ (Fig. 2a) confirms the scaling
theory by Affleck and Halperin for this problem\cite{AfflH}. 
The behavior of the spin-spin correlator for an isolated chain $
\langle S(x) S(0) \rangle \sim (1 / x) \exp ( - x / \xi_1)$, 
$\xi_1 \sim 1/T$ signals
the approach to the Gaussian fixed point: within the thermal length $\xi_1$
the system exhibits algebraic long range order. For finite $\alpha$ the
crossover temperature $T^0$ can be found using the standard mean-field
consideration: at $T^0$ temperature becomes of order of the exchange
interaction between two patches of correlated spin on neighboring chains
of size $\xi_1$: $k_B T^0 \simeq
\alpha \Phi_1^2 / \xi_1 (T^0)$, where $\Phi_1 = \xi_1 (T^0) \phi$,
$\phi$ being the microscopic staggered magnetization. Taking $\phi$ independent
of $\alpha$ would yield the erreneous result that $T^0 \sim \sqrt{\alpha}$.
The subtlety is that when $\alpha$ is sufficiently small, the quantum dynamics 
within the correlation volume $\xi_1$ is already in the 2+1D
regime\cite{vDZa}. Using the $T=0$ result by Affleck and Halperin that
$\phi \sim \sqrt{\alpha}$\cite{AfflH},
 we recover $T^0 \sim \alpha$, $\alpha << 1$.
 The other feature
worthwhile mentioning is that $T^0$ and $T^*$ are identical for the 1- and 3-leg cases for small $\alpha$'s. This is in line with the observations by 
Frishmut {\em et. al.}\cite{FAT} 
that these spin ladders renormalize in identical Luttinger liquids when
the ladder exchange interactions are isotropic.

In the two-leg ladders case (Fig. 2b) the quantum
order-disorder transition occurs at a finite value of $\alpha$, 
$\alpha_c = 0.30(2)$. This is in line with the qualitative expectations
(see also \cite{GRS,ADB}). Since the isolated two leg ladders are 
incompressible spin systems, the ladder-to ladder interaction has to overcome
the single ladder energy gap before the two dimensional lock-in can occur.
This critical $\alpha$ is rather large, and in
addition, the 1+1 D $\rightarrow$ 2+1 D crossover temperature $T^0$ shows
the upward curvature ($T^0 \sim \sqrt{\alpha}$) previously predicted from
a scaling analysis of the anisotropic QNLS model (AQNLS)\cite{vDZa}. 
As a ramification, $T^0$ and $T^*$ (as well as $T'$) separate
and a {\em large, genuinely 2+1 D quantum critical regime opens up} around
$\alpha_c$. This is in marked contrast with the isotropic Heisenberg model where
the renormalized classical regime sets in essentially at the 
lattice cut-off\cite{SaSc,MaDi}.

The gross $\alpha$ dependences of the various cross-over temperatures can be
understood by considering the AQNLS model obtained by taking the naive
continuum limit for the ladder problem.
An average staggered field $\vec{\phi}$ is introduced for a block of 
$2\times n_l$ sites. Integrating out the quadratic fluctuations\cite{Se}, the
effective action for $\vec{\phi}$ becomes the AQNLS model with anisotropic 
spin wave velocities,
\begin{eqnarray}
c^2_x&=& \alpha c^2_0 
\left\{ 
       \begin{array}{cl}
       \frac{(3+\alpha)}{2(1+\alpha)} & {\rm for~} n_l=2\\
       \frac{9(7+3\alpha)}{2(1+2\alpha)(13+2\alpha)} & {\rm for~} n_l=3
       \end{array}
\right. \\
c^2_y&=&c^2_0 
\left\{ 
       \begin{array}{cl}
       \frac{(3+\alpha)}{4} & {\rm for~} n_l=2\\
       \frac{3(7+3\alpha)}{2(13+2\alpha)} & {\rm for~} n_l=3
       \end{array}
\right. ,
\label{spwa}
\end{eqnarray}
where $c_0$ is the spin wave velocity in the isotropic limit. The
coupling constant $g_0$ is $\alpha$ independent and the same 
as for the isotropic model. According to the scaling analysis of 
Ref.\cite{vDZa}, the renormalized spin-stiffness becomes in terms 
of the velocities $c_{x,y}$,
\begin{equation}
\rho_s(\alpha)=\rho_s\frac{c_x(1-\frac{g_0}{g_c(\alpha)})}
{c_y(1-\frac{g_0}{g_c(1)}) }
\end{equation}
where
\begin{eqnarray}
g_c(\alpha)=&&4\pi\sqrt{c_0/c_y} \left(1+\frac{2}{\pi}(c_y {\rm arsinh}[c_x/c_y]/c_x \right.\nonumber\\
&&\left.+{\rm ln}[c_y(1+\sqrt{1+c_x^2/c_y^2})/c_x/(1+\sqrt{2})^2])\right)
\end{eqnarray}
and $\rho_s$ is the spin stiffness for $\alpha=1$. According to Ref.\cite{vDZa},
the cross-over scales are $T^* = 2\pi \rho_s(\alpha)$, 
$T^0 = 2\pi \rho_s c_x (g_0/(4\pi c_0)+(1-g_0/g_c)/c_y)$
and $T' = {\rm const.} |\rho_s(\alpha)|$. It 
turns out that for the bare coupling constant $g^0$ as determined for the
isotropic case ($g^0 = 9.1$), the order-disorder transition occurs 
at a  somewhat small value of $\alpha= 0.08$, which is not 
surprising given the approximations involved (one-loop level). 
However, by adjusting $g_0$ to shift $\alpha_c$ to its numerical
value ($g_0 = 11.0$), we find a very close
agreement between the numerical- and analytical results
for the various cross-over temperatures (Fig. 2b). As can be seen
from Fig. (2a), the above analysis also works quite well for the
three-leg ladders for $\alpha \geq 0.4$. Remarkably, it
seems that $T^*$ switches rather suddenly from the AQNLS behavior
at large $\alpha$ to the linear behavior expected for the Luttinger
liquid regime, as if the topological terms start to dominate rather
suddenly. 

Besides its intrinsic interest, the above does have potentially important
ramifications for the understanding of the quantum-magnetism in cuprates:
bond ordering of stripes would imply that already
at rather moderate values of the anisotropy $\alpha$,  spin-ladder
physics alone would enhance the quantum spin fluctuations substantially. 
This can be further illustrated by comparing the temperature dependence of
$T \xi_y (T)$ for the isotropic spin system $\alpha = 1$ with that of 
the coupled two-leg ladders in the vicinity of the critical $\alpha$
(Fig. 3). This quantity can be directly compared with the spin-spin
relaxation rate $1/T_{2G}$ and, with some caution, also to $1/T_1$\cite{SaSc,SDS} 
(a dynamical critical exponent $z=1$ is only strictly obeyed in the QC regime).
As compared to the isotropic case, the exponential increase of $T \xi$
(signalling the renormalized classical regime) is shifted to a low temperature,
while over most of the temperature range $T \xi(T)$ is constant, as is
found in cuprates. It is noted that the `quantum-critical
signature' $\xi \sim 1/T$ extents in the temperature range above the dimensional
crossover temperature $T^0$. Since this regime is non-universal this
should be regarded as a quasi-criticality.

\begin{figure}
\vspace{-7\baselineskip}
\epsfig{figure=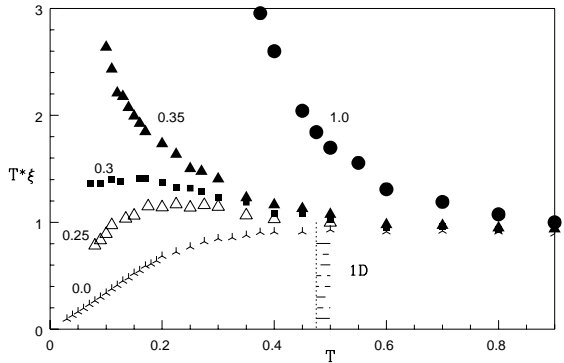, width=8cm}
\caption{ $\xi_y T$ versus temperature
for the 2-leg system, when the $\alpha$'s 
are close to critical point. Results for
$\alpha=0.0$ (isolated ladders) and $1.0$ (isotropic limit)
are added for comparison. The vertical bar indicates the 1D-2D cross-over
temperature.}
\end{figure}
       
\vspace{-\baselineskip}
This is no more than suggestive. However, it points at a simple strategy to
clear up these matters by experiments involving the static stripe phase. 
It should be established if the stripe phase is site- or bond ordered which
can be done by NMR. Next, the $\alpha$ should be determined 
from neutron measurements of the spin-wave velocities, Eq.(\ref{spwa}). 
Using these
as an input, the temperature dependence of the correlation-length, as well
as the NMR relaxation rates, can be calculated to a high precision 
starting from a microscopic spin-only dynamics. Comparison of these
quantities to experiment should yield insights in the microscopic origin
of the peculiar spin dynamics in doped cuprates.

{\em Acknowledgments.} 
We thank B. I. Halperin for helpful discussions. 
 J.T. acknowledges a
fellowship supported by Foundation for Polish Science (FNP), and J.Z. 
financial support by the Dutch Academy of Sciences (KNAW).

\end{multicols}

\end{document}